\documentclass[12pt,preprint]{aastex}
\usepackage{emulateapj5,mathptmx}

\submitted{Accepted for Publication in The Astrophysical Journal}

\makeatletter

\newenvironment{inlinefigure}{%
\def\@captype{figure}%
\noindent\begin{minipage}{0.999\linewidth}\begin{center}}
{\end{center}\end{minipage}\smallskip}
\makeatother

\newcommand{\gsim}{\mbox{\hspace{.2em}\raisebox{.5ex}{$>$}\hspace{-.8em}\raisebox{-.5ex}{$\sim$}\hspace{.2em}}}  
\newcommand{\lsim}{\mbox{\hspace{.2em}\raisebox{.5ex}{$<$}\hspace{-.8em}\raisebox{-.5ex}{$\sim$}\hspace{.2em}}}  
\newcommand{\ssst}{\scriptscriptstyle}
\newcommand{\E}[1]{\times 10^{#1}}
\newcommand{\RA}[3]{\mbox{R.A.}={#1}^{{\rm h}}{#2}^{{\rm m}}{#3}^{{\rm s}}}
\newcommand{\decl}[3]{\mbox{decl.}={#1}^{\circ}{#2}'{#3}''}

\newcommand{\s}{\,{\rm s}} 	\newcommand{\ps}{\,{\rm s}^{-1}}
\newcommand{\yr}{\,{\rm yr}}	
\newcommand{\cm}{\,{\rm cm}}	\newcommand{\km}{\,{\rm km}}
\newcommand{\kmps}{\km\s^{-1}}
\newcommand{\parsec}{\,{\rm pc}}\newcommand{\kpc}{\,{\rm kpc}}
\newcommand{\ergs}{\,{\rm erg}}	\newcommand{\K}{\,{\rm K}}
	\newcommand{\Msun}{M_{\odot}}
	\newcommand{\keV}{\,{\rm keV}}
\newcommand{\G}{\,{\rm G}}

\newcommand{\nel}{n_{e}}	\newcommand{\NH}{N_{\ssst\rm H}}
	
\newcommand{\Ts}{T_{s}}		\newcommand{\vs}{v_{s}}
\newcommand{\nH}{n_{\ssst\rm H}} 	\newcommand{\mH}{m_{\ssst\rm H}}
\newcommand{\xray}{X-ray}

\newcommand{\rosat}{{\em ROSAT}} \newcommand{\asca}{{\em ASCA}}
\newcommand{\du}{d_{22}}

\newcommand{\oh}{OH(1720 MHz)}
\newcommand{\snr}{G349.7+0.2}

\begin{document}

\title{An ASCA Study of the High Luminosity SNR G349.7+0.2}

\author{Patrick Slane\altaffilmark{1}, Yang Chen\altaffilmark{2},
Jasmina S. Lazendic\altaffilmark{1}, and John P. Hughes\altaffilmark{3}}

\altaffiltext{1}{Harvard-Smithsonian Center for Astrophysics,
    60 Garden Street, Cambridge, MA 02138}
\altaffiltext{2}{Department of Astronomy, Nanjing University, Nanjing
210093, P. R. China}
\altaffiltext{3}{Department of Physics and Astronomy, Rutgers, The State
University of New Jersey, Piscataway, NJ 08854-8019}

\begin{abstract}

We present {\em ASCA} observations of supernova remnant (SNR)
\snr. The remnant has an irregular shell morphology and is
interacting with a molecular cloud, evident from the presence of \oh\
masers and shocked molecular gas. The X-ray morphology is consistent 
with that at radio
wavelengths, with a distinct enhancement in the south. The X-ray emission from
the SNR is well described by a model of a thermal plasma which has yet to reach
ionization equilibrium. 
The hydrogen column of $\sim 6.0 \times 10^{22}$ cm$^{-2}$ is consistent with
the large distance to the remnant of $\sim$ 22 kpc estimated from the
maser velocities. We derive an X-ray luminosity of $L_{x}(0.5$-$10.0\keV)=
1.8\E{37}\du^{2}\ergs\ps$, which makes
G349.7+0.2 one of the most \xray\ luminous shell-type SNRs known in the Galaxy. 
The age of the
remnant is estimated to be $\sim$ 2800 yrs. The ambient density and pressure
conditions appear similar to those inferred for luminous compact SNRs found
in starburst regions of other galaxies, and provides support for the notion
that these may be the result of SNR evolution in the vicinity of dense
molecular clouds.

\end{abstract}


\section{Introduction}
\label{sec:intro}

Because massive stars evolve quickly, they are often not far from
their birth sites when they expire. The result is that many of the
supernova remnants (SNRs) produced in the explosive events that mark
the endpoint of stellar evolution for these stars are located near
the molecular cloud complexes from which the progenitors emerged. The
initial expansion of such an SNR is likely to proceed rather
effortlessly as the progenitor star has generally sculpted a cavity in
the ambient medium by virtue of a strong wind (Chevalier 1999). 
Eventually the blast
wave must contend with the cavity walls, however, and when the cavity
resides in a dense molecular cloud the resulting interaction reveals
itself spectacularly in X-rays (Chevalier \& Liang 1989). The remnant 
sweeps up massive amounts
of material and heats it to X-ray emitting temperatures while seeding
the cloud with metals synthesized in the supernova explosion. 
Such young SNRs encountering dense material can transform a large
amount of their kinetic energy into radiation, appearing as
bright (radio and \xray) emission sources, often with irregular
morphologies.
They may be representative of a larger class of compact SNRs identified
as bright radio sources in starburst regions of other galaxies (e.g.
Kronberg, Biermann, \& Schwab 1985; Antonucci \& Ulvestad 1988; and Smith
et al. 1998). Chevalier \& Fransson (2001) have proposed that these sources
represent SNRs that have evolved in the high density interclump medium
of molecular clouds, and that a similar population that has escaped such
high density regions is responsible for driving galactic winds in the
host galaxies.

A good example of this type of SNR is N132D in the Large Magellanic
Cloud (LMC). It is a luminous ($\sim 5 \E{37}\ergs\s^{-1}$ in X-ray
band) small diameter SNR ($\sim$ 44\arcsec = 11.7 pc) that is evolving into a
cavity wall on the edge of a molecular cloud
(Hughes 1987, Banas et al. 1997).
The X-ray emitting material comprises several
hundred solar masses and the overall abundances are characteristic of
the LMC interstellar material,  implying that the bulk of the emission
is from swept--up material. However, optical spectra (Danziger \& 
Dennefeld 1976) show that N132D is an oxygen rich SNR whose abundances
are consistent with a $\sim 20 M_\odot$ progenitor (Blair, Raymond, \&
Long 1994), and high resolution X-ray spectral studies
show the presence of an ejecta component as well 
(Hwang et al. 1993, Behar et al. 2001)
indicating that N132D is a relatively young SNR; the large amount of 
swept-up material, as well as the high X-ray luminosity, are the result
of an explosion in the high density surroundings of a molecular cloud. 

G349.7+0.2 appears in some ways to be a Galactic counterpart to N132D. It is an
SNR with a small angular size ($r \sim 1$~arcmin) and the third
highest radio surface brightness next to Cas~A and the Crab
Nebula. Its nonthermal radio emission ($\alpha_{r}\simeq-0.5$) and roughly circular morphology, classifies it as a shell--type SNR
(Shaver et al. 1985).
However, it has an emission peak near the
southeastern edge rather than a prominent limb--brightened
structure and central cavity typical of shell--type remnants.
The early H\,{\sc i} absorption measurements showed that 
G349.7+0.2 lies beyond the tangent point, with the kinematic distance
in the range $13.7<d<23\kpc$  (Caswell et al. 1975).
\oh\ masers have been
recently found interior to G349.7+0.2 with radial velocities 
$\sim +16 {\rm\ km\s}^{-1}$ indicating a kinematic distance 
$d \approx 22.4$~kpc (Frail et al. 1996), for which the SNR radius is 6.4~kpc.
This maser emission, detected in about 10\% of the Galactic
SNRs, has been recognized as a signpost of SNR interactions with
molecular clouds (see Koralesky et al. 1998 and references therein; also
Lockett, Gauthier \& Elitzur 1999).
The masers are located along
the radio ridge extending from the southern emission peak to the
north. CO observations toward G349.7+0.2 revealed a molecular cloud
associated with the SNR, and with the masers (Reynoso \& Mangum 2000,
Reynoso \& Mangum 2001) which
have been produced by shocks from the expanding SNR blast wave 
(Lazendic et al. 2002).

\begin{inlinefigure}
  \includegraphics[width=1.2\linewidth]{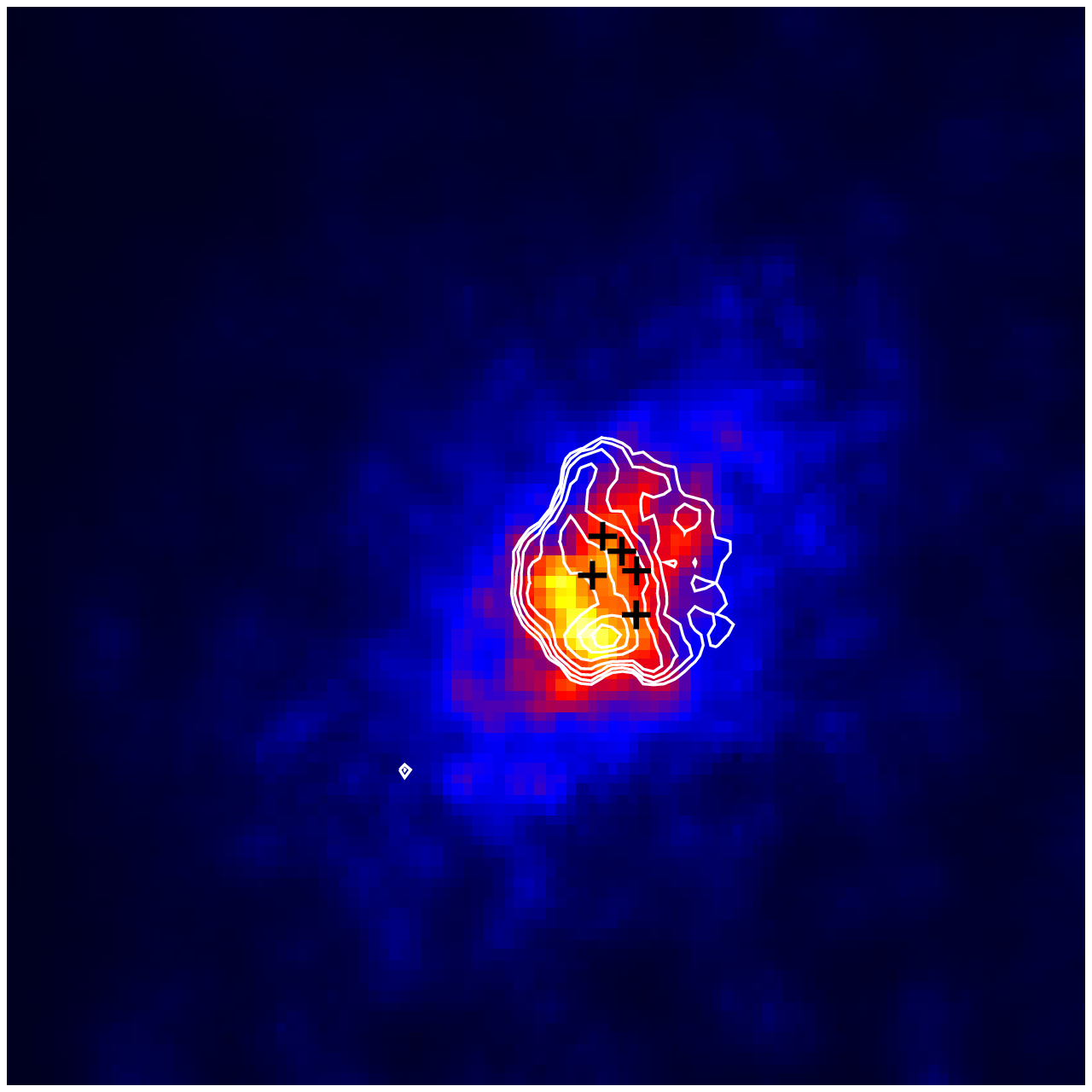}
  \caption{
  The broad band (1.0-10.0 keV) {\em ASCA} SIS image of \snr.
  Contours are from the ATCA 18--cm continuum image,
  with levels at 10, 17, 33, 66, 132, 198 and 264 mJ\,beam$^{-1}$.
  Crosses mark \oh\ maser positions.
  Comparison of \emph{Chandra} and \emph{ASCA}
  }
\vspace{0.1in}
\label{fig-asca}
\end{inlinefigure}

G349.7+0.2 went undetected in the \rosat\ all--sky survey. As we show
below, this is the result of the high interstellar absorption of soft
\xray\ photons. The remnant was detected in the \asca\ Galactic plane
survey (Yamauchi et al. 1998),
and we have followed up this detection with
a 60 ks observation that was obtained in two pointings.

\section{ASCA Data}

The pointed observations of G349.7+0.2 were carried out with \asca\
satellite on 16--17 March 1998  (25 ks) and 19--20 March 1999 (36 ks)
with the solid--state imaging spectrometer (SIS) and the gas imaging
spectrometer (GIS).  The SIS observations were performed in 1--CCD
mode to provide the best spectral resolution. The single-chip field
of view provides sufficient sky coverage for independent background
subtraction.
The data were screened according to standard procedures for GIS and 
SIS data, resulting in roughly 48(53)~ks of good exposure time 
with $\sim 6500(12\,000)$ SIS(GIS) events from the SNR in each camera.

\subsection{X-ray Image of \snr}

In Figure~\ref{fig-asca}
we present the exposure-corrected 1--10~keV X-ray image of \snr\ 
produced by merging the SIS0 and SIS1 maps. Each map from the two
exposures was adaptively smoothed using a Gaussian 
kernel with a minimum of 50 counts in each smoothing element. 
Inspection of images in different energy bands show no
obvious variations in morphology. However, the modest angular
resolution of {\em ASCA} limits our ability to detect variations on
scales below $\sim$2\arcmin. The radio image, shown as contours in
Figure~\ref{fig-asca}, is characterized by a distinct 
brightness enhancement along the southwestern limb, where the remnant
is interacting with a molecular cloud. The X-ray brightness
is enhanced in this region as well, and the overall morphology is
similar to that observed in the radio. The larger apparent size in X-rays,
with emission extending beyond the radio shell,
is largely an artifact of the broad angular response of the X-ray telescope.
In the analysis presented below we use the radio size of the SNR to 
estimate its radius. Higher resolution X-ray observations will be required
to fully explore the distribution of the X-ray emitting gas.

\subsection{Spectral Analysis}
To investigate the spectrum of \snr, we used both GIS and SIS data. 
For each GIS detector (GIS2 and GIS3), data from the two observations were 
merged to form a single spectrum. Based on the centroids of the spectral
lines, a gain adjustment of $\sim$3\% was required, 
consistent with known uncertainties in the GIS calibration.
The spectra were extracted from a circular region with a radius
6\farcm 8, centered at $\RA{17}{18}{03}$,  $\decl{-37}{26}{07}$
(J2000). The spectra from the four SIS data sets were
extracted from a region with the same center, but with a radius of
2\farcm 8. For both the SIS and GIS, background spectra were extracted
from the outer regions of  the detectors. Spectra were re--grouped to
include at least 25 net counts per bin. Because of resolution changes between
the two observations, data from the same cameras (SIS0 and SIS1) were not
combined. Our spectral analyses were thus based on simultaneous fitting
of two GIS spectra and four SIS spectra.

\begin{inlinefigure}
  \includegraphics[width=0.85\linewidth]{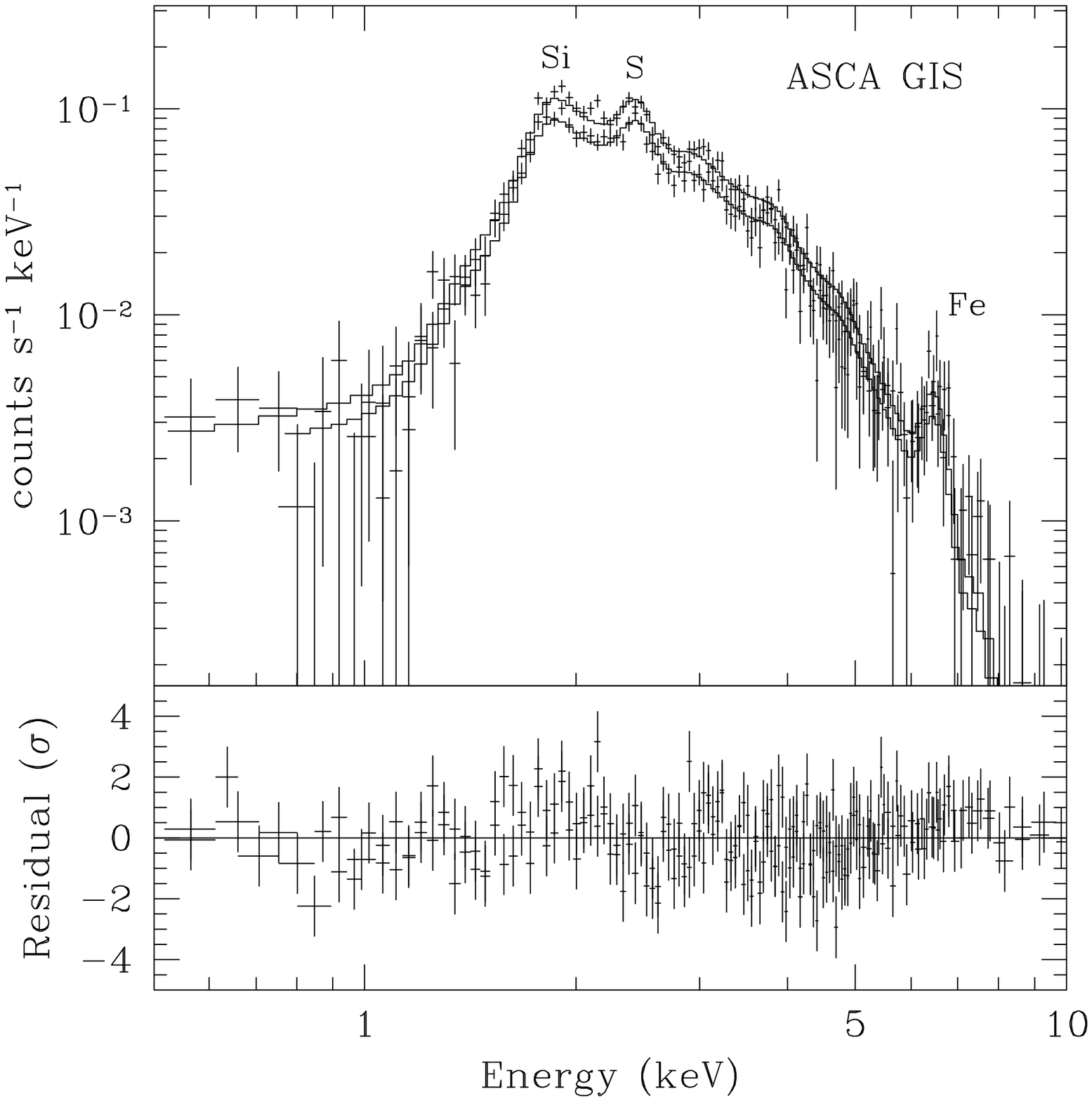}
  \caption{
  ASCA GIS spectra and residuals for the NEI fit
  described in the text.
  }
\label{fig-spectrum}
\vspace{0.1in}
\end{inlinefigure}

\begin{table*}
\caption{
  Results from CIE model (Raymond \& Smith 1977) fit to 2 GIS and 4
  SIS data sets, with the 90\% confidence ranges.\label{tab-fit}}
\vspace{-0.1in}
\begin{center}
\begin{tabular}{cccc}
\hline\hline
& \multicolumn{3}{c}{Value}\\
Parameter               & CEI & NEI & VPSHOCK\\
\hline
$f\nel\nH V/(d/22\kpc)^{2}$ ($10^{59}\cm^{-3}$) & $8.8$
7.0 & 10.2\\
$T_{x}$ (keV)   & $1.05^{+0.02}_{-0.03}$ & $1.26^{+0.03}_{-0.05}$ & $1.12 \pm
0.03$ \\
$\NH$ ($10^{22}\cm^{-2}$) & $5.7\pm0.1$ & $5.8 \pm 0.1$ & $5.9 \pm 0.1$  \\
$n_e t (10^{11} \rm\ cm^{-3}{\rm\ s}$) & & $1.7^{+0.5}_{-0.3}$ &
$6.5^{+2.9}_{-1.8}$$^a$ \\
$F(0.5$-$10\keV)$ ($\ergs\cm^{-2}\ps$) & $7.0\E{-12}$ & $7.1\E{-12}$
& $7.2\E{-12}$ \\
$F_{\rm unabs}(0.5$-$10\keV)$ ($\ergs\cm^{-2}\ps$) & $1.4\E{-10}$ &
$1.7\E{-10}$ & $2.8\E{-10}$ \\
$\chi^{2}/{\rm d.o.f.}$ & $708.5/599$ & $669.6/598$ &$661.0/598$ \\
\hline
\end{tabular}\\
a) Represents maximum value of continuous range in {\tt VPSHOCK} model\\
\end{center}
\vspace{-0.2in}
\end{table*}

As illustrated in Figure~\ref{fig-spectrum}, the spectrum is dominated
by line features from Si, S, and Fe-K. There is no 
emission below $\sim1.5\keV$, indicating a
large value of line--of--sight hydrogen column density $\NH$.
The centroids of the distinct line features were determined by
fitting Gaussian profiles to these regions of the spectra. We obtain 
$1.86^{+0.02}_{-0.01}\keV$ for Si~He$\alpha$,
$2.42^{+0.02}_{-0.01}\keV$ for S~He$\alpha$, and
$6.53^{+0.09}_{-0.10}\keV$ for Fe~K$\alpha$.
The broadband spectrum can be modeled adequately ($\chi_\nu^2=1.18$) 
as an optically thin thermal plasma ($kT_x = 1.05 \pm 0.03$~keV)
in collisional ionization equilibrium (CIE; 
Raymond \& Smith 1977)
modified by interstellar absorption 
(Morrison \& McCammon 1983) of
$N_H = (5.7 \pm 0.1) \times 10^{22}{\rm\ cm}^{-2}$, 
and with normal solar abundances\footnote{Uncertainties here and throughout the
paper represent 90\% confidence intervals}. This indicates that 
the X-ray emitting
plasma of G349.7+0.2 is dominated by the swept--up interstellar
material, with no significant evidence of emission from the supernova ejecta.

An improved fit  ($\Delta \chi^2 = 39.1$ for one additional fit 
parameter) is obtained with a single-temperature non--equilibrium
ionization (NEI) model 
(Hughes \& Singh 1994)
with 
$kT = 1.26 \pm 0.05$~keV and an ionization timescale $n_e t = 
(1.7 \pm 0.4) \times 10^{11} {\rm\ s\ cm}^{-3}$, indicating 
that the plasma is still ionizing. A single-temperature plane-parallel
shock model ({\tt VPSHOCK}, Borkowski et al. 2001)
yields an equally 
good fit with similar
parameters. Here the ionization parameter is allowed to vary from zero
to a fit-determined maximum value of $6.5^{2.9}_{1.8} \times 10^{11} 
{\rm\ s\ cm}^{-3}$; this range is consistent with the average value 
determined in the single ionization timescale model. 
No significant contribution of a power law component is found.
The spectral fit results are tabulated in Table~\ref{tab-fit}.

The hydrogen column $\NH\sim6\E{22}\cm^{-2}$  confirms a substantial
absorption of the soft emission ($<1.5\keV$) and is consistent with a
very large distance. The corresponding \xray\ luminosity (using the NEI fit)
is $L_{x}(0.5$-$10\keV)= 1.8\E{37}\du^{2}\ergs\ps$, which makes
G349.7+0.2 one of the most \xray\ luminous Galactic SNRs known, rivaling
even the young ejecta-dominated remnant Cas~A.
The radius for \snr\ is $\sim 1$~arcmin, for which its dimensional size is
$r_{s}\sim6.4\du\parsec$. The volume emission measure 
produces a mean hydrogen density
$\nH\sim4.2f^{-1/2}\du^{-1/2}\cm^{-3}$ (assuming $\nel\approx1.2\nH$),
where $f$ is the volume filling factor (i.e. $V = \frac{4}{3}\pi R^3 f$
is the volume of the X-ray emitting region, where $R$ is the SNR radius);
for a Sedov-phase SNR, $f = 0.25$ (Sedov 1959).
The \xray\ emitting mass 
$M_x=1.4\nH \mH f V$ is $\sim160f^{1/2}\du^{5/2}\Msun$.

Assuming \snr\ is in the adiabatic phase of evolution
(Sedov 1959),
the post--shock temperature is
$kT_{s}\approx0.77kT_{x}\sim 1\keV$ ($T_{x}\approx1.2\E{7}\K$).
The blast wave velocity is accordingly
$\vs=(16k\Ts/3\mu\mH)^{1/2}\sim900\kmps$ (where the mean atomic
weight $\mu=0.604$). Thus, an estimate of the age of the remnant is
$t=2r_{s}/5\vs\sim2800\du\yr$. This estimate results directly from
the X-ray temperature fit and should be more reliable than
the $1.4\E{4}\yr$ 
Reynoso \& Mangum (2001)
obtained using an assumed
explosion energy and adopting a rough estimate of the cloud density as 
the remnant's ambient density. The explosion energy is approximately
$E=1.12\times10^{-24}n_0 (r_{s}^{5}/t^{2})\sim5\E{50}f^{-1/2}\du^{5/2}\ergs$.
We note, however, that the electron temperature in the shocked gas (i.e. the
temperature inferred from X-ray spectral fits) lags the ion temperature,
while it is the latter that determines the SNR dynamics. If full temperature
equilibration has not yet occurred in \snr, 
so that the shock velocity is actually higher than that indicated by the
electron temperature, then the age given above is an
overestimate while the explosion energy is an underestimate.

\section{Discussion}

The normal abundances and large X-ray emitting mass found in
\snr\  indicate that the gas is dominated by swept--up interstellar
material, which is consistent with the SNR location near the molecular
cloud. 
The maximum radio and X-ray brightness occurs in a region
where the SNR shell is encountering a
molecular cloud with a radius of 24\arcsec\ (2.6 pc), a mass of
1.2$\times10^{4}\Msun$, and a gas density of $\sim10^{4}\cm^{-3}$
(Reynoso \& Mangum 2001).
This cloud, as well as \snr, lies along a large CO filament 
(Figure~\ref{fig-atca+co}).
Beyond the positional coincidence,  an on--going
interaction between \snr\ and  the molecular  cloud is implied by
the presence of \oh\  masers  and the detection of  shocked molecular
gas. The \oh\ maser spots  detected from G349.7+0.2  
(Frail et al. 1996)
are attributed to C--shock collisional  excitation in the clumpy
molecular clouds, with  cloud temperature  and density conditions $50\le T\le125\K$ and $n_{{\ssst\rm  H}_{2}}\sim10^{5}\cm^{-3}$
(Lockett, Gauthier \& Elitzur 1999).
Molecular--line observations toward the
maser region in \snr\ confirm the presence of shocked molecular gas
with density $10^5-10^6\cm^{-3}$ and temperature $>$ 40 K  
(Lazendic et al. 2002).

\begin{inlinefigure}
\vspace{0.05in}
  \includegraphics[width=1.0\linewidth]{f3.ps}
  \caption{
  Velocity--integrated CO 1-0 image of the region around
  \snr\ from
  Reynoso \& Mangum (2001)
  overlaid with the radio contours. The
  greyscale range is from $-$6 K to 51 K\,km\,s$^{-1}$. The
  contour levels are same as in Fig.~\ref{fig-asca}.}
\label{fig-atca+co}
\end{inlinefigure}

The adjacency of \snr\ to dense molecular clouds suggests that its
progenitor was a massive star that did not move significantly from
its birthplace prior to explosion.
Early type (O4 -- B0) massive stars, during their
main sequence lifetime, will photoevaporate nearby clouds and
homogenize a region of radius $R_{h}=56n_{m}^{-0.3}\parsec$, where
$n_{m}$ denotes the mean medium density 
(McKee, van Buren, \& Lazareff 1984).
Since the
maser clouds exist at $r\lsim 6.4\du\parsec$, the medium density
should satisfy $n_{m}\gsim1.4\E{3}\du^{3.3}\cm^{-3}$ if the progenitor
was earlier than B0. This mean density is consistent with that expected in the
molecular clouds which harbor \oh\ masers. 

The thermal pressure of the hot X-ray shell is 
$P_{\rm sh} \approx 2.3 n_H k T_s \sim 1.6\E{-8}f^{-1/2} d_{22}^{-1/2}$
${\rm erg\ cm}^{-3}$. This is higher than the thermal gas pressure in the maser
portions of the shocked clumps, where $P_{\rm cl} = n_{{\ssst\rm H}_{2}} k T
\sim 9 \times 10^{-9} \ergs\cm^{-3}$, using the upper range of 
$n_{{\ssst\rm H}_{2}}$ and $kT$ allowable for the production of
strong masers (see above). 
Moreover, if electron-ion equilibration has not yet been achieved in
the X-ray shell, the pressure in this gas is even higher. The
discrepancy implies that there must be a contribution from the
magnetic pressure and/or cosmic ray pressure in the shocked
molecular gas. The strength of the line--of--sight magnetic field
determined using Zeeman splitting of the \oh\
line is $B_{\theta} = 3.5 \pm 0.5 \E{-4}\G$ 
(Brogan et al. 2000), where $\theta$ is
the angle between the magnetic field vector and the line of sight.  
Similar measurements in other remnants have demonstrated that the
$B_{\theta}$ values are not confined just to the maser locations but
represent a global measurement of the magnetic field in the post--shock
molecular gas (Claussen et al. 1997).
This yields a magnetic pressure $P_{\rm mag} = B_{\theta}^2/8\pi 
\approx 5 \times 10^{-9}$ erg\,cm$^{-3}$ which still falls short of 
the pressure in the shell. Further,
if the masers are saturated, the true maser field may be as much as a factor 
of 5 lower than that inferred from Zeeman splitting 
(Brogan et al. 2000).  This would appear to suggest
either a slight overpressure in the SNR shell or a non-negligible
cosmic ray pressure component in the molecular gas. The latter would appear
plausible in that the radiative shock in the molecular gas will
suffer rapid cooling and compression, resulting in an increase in the density
and cosmic ray pressure. 

The overall properties of \snr\ are similar to those proposed by 
Chevalier \& Fransson (2001) to explain the nature of compact SNRs
in starburst galaxies such as M82. The high ambient density associated
with the presence of molecular clouds leads to rapid evolution of the SNR
and a short-lived period of very high luminosity,
after which the shell becomes radiative and the SNR cools rapidly. 
The radio emission from
such remnants persists for longer than the X-ray emission, which is consistent
with the fact that those in starburst galaxies are primarily recognized based
on their radio luminosity. \snr\ is still dynamically young enough to produce a
significant flux of X-rays as well. The density, age, and luminosity inferred 
above are in good agreement with the model predictions of Chevalier \& Fransson
(2001); \snr\ appears to be several thousand years old
and has evolved in the vicinity of a complex of molecular clouds,
apparently encountering gas with an average density of order $5 {\rm\
cm}^{-3}$, and driving a shock into more dense adjacent molecular gas, thereby 
exciting OH maser emission. If this scenario is indeed correct, then the region
in which the SNR shell has gone radiative should be accompanied by a an
HI shell. High resolution mapping of this region in HI is this of considerable
interest.

\section{Conclusions}

The \asca\ SIS and GIS spectra of the SNR G349.7+0.2
are well described by a model for a somewhat underionized, optically thin,
thermal plasma.
The spectral analysis confirms that it is a distant,
young SNR that has swept up considerable interstellar material, consistent
with the picture that \snr\ is interacting with an adjacent molecular cloud
that is observed in CO. Velocity measures based on OH masers produced by 
this cloud interaction yield a distance to the remnant, based upon which we
conclude that \snr\ is one of the most X-ray luminous shell-type SNRs known 
in the Galaxy. Combined with its high radio luminosity, this \snr\ appears to 
be a young analog of the bright SNRs observed in starburst galaxies, many of
which have evolved in such high density/pressure regions that they have already
become radiative and X-ray faint. We note that N132D, a similar SNR in the
LMC, shows evidence for ejecta in its X-ray emission. Given the
relatively young age, it may be that such a component persists in
G349.7+0.2 as well, but is dominated by the emission from swept up
material. The high column density to the SNR rules out any opportunity to
detect oxygen ejecta, but evidence for Fe-K ejecta is also seen in N132D, 
and could be observable in G349.7+0.2 with higher sensitivity measurements.
Future high resolution X-ray studies, already planned, offer the possibility 
to search for such a component. 

\acknowledgements{We thank Dick Edgar and John Raymond for helpful
discussions related to this study, as well as the referee, Rob Petre, for
his careful review of the text. Y.C. acknowledges support from CNSF 
grant 1007003 and grant NKBRSF-G19990754 of China Ministry of Science 
and Technology. This work was also supported, in part, by NASA 
contract NAS8-39073 and grant NAG5-8354 (POS).}

\end{document}